\documentclass[aps,prl,twocolumn,showpacs,superscriptaddress,10pt]{revtex4-1}
\usepackage{bbm}
\usepackage{graphicx}
\usepackage{amsmath}
\usepackage{amssymb,amsthm}
\usepackage{hyperref}
\usepackage[normalem]{ulem}
\usepackage{color}
\usepackage{mathtools}

\newcommand{\ket}[1]{\vert #1 \rangle}

\newcommand{\eq}{\begin{eqnarray}} 
\newcommand{\en}{\end{eqnarray}}

\begin{document}

\title{Capturing the Feshbach-induced Pairing Physics in the BEC-BCS Crossover}

\author{Eloisa Cuestas}
\affiliation{Universidad Nacional de C\'ordoba, Facultad de Matem\'atica, Astronom\'ia, F\'isica y Computaci\'on, Av. Medina Allende s/n, Ciudad Universitaria, X5000HUA C\'ordoba, Argentina}
\affiliation{Instituto de F\'isica Enrique Gaviola (IFEG), Consejo de Investigaciones Cient\'ificas y T\'ecnicas de la Rep\'ublica Argentina (CONICET), C\'ordoba, Argentina}

\author{Jos\'e I. Robledo}
\affiliation{Centro Atomico Bariloche (CAB), Consejo de Investigaciones Cient\'ificas y T\'ecnicas de la Rep\'ublica Argentina (CONICET), Bariloche, Argentina}

\author{Ana P. Majtey}
\affiliation{Universidad Nacional de C\'ordoba, Facultad de Matem\'atica, Astronom\'ia, F\'isica y Computaci\'on, Av. Medina Allende s/n, Ciudad Universitaria, X5000HUA C\'ordoba, Argentina}
\affiliation{Instituto de F\'isica Enrique Gaviola (IFEG), Consejo de Investigaciones Cient\'ificas y T\'ecnicas de la Rep\'ublica Argentina (CONICET), C\'ordoba, Argentina}

\date{\today}

%%%%%%%%%%%%%%%%
%%%%%%%%%%%%%%%%

\begin{abstract}
By including the effect of a trap with characteristic energy given by the Fermi temperature $T_F$ in a two-body two-channel model for Feshbach resonances, we reproduce the experimental closed-channel fraction $Z$ across the BEC-BCS crossover and into the BCS regime of a $^6$Li atomic Fermi gas. We obtain the expected behavior $Z \propto \sqrt{T_F}$ at unitarity, together with the recently measured proportionality constant. Our results are also in agreement with recent measurements of the $Z$ dependency on $T_F$ on the BCS side, where a significant discrepancy between experiments and theory has been repeatedly reported.
\end{abstract}

\maketitle

%%%%%%%%%%%%%%%%
%%%%%%%%%%%%%%%%
% Introduction
%%%%%%%%%%%%%%%%
%%%%%%%%%%%%%%%%

Magnetic-field tunable Feshbach resonances provide the essential tool to control the interaction between atoms in ultracold quantum gases. In current ultracold gases experiments these resonances are induced by varying the strength of an external magnetic field used to tune the relative energy between the collision energy of two atoms and that of a quasibound molecular state via the Zeeman effect. The resonant interactions allow not only to control the strength of the atomic interactions but also if they are effectively repulsive or attractive \cite{regal_2003_PRL, chin_2010_review}. Over the last twenty years, this precise generation and control of interactions has been a crucial ingredient in the understanding of the behavior of quantum matter, leading to many breakthroughs such as the generation of fermionic Bose Einstein condensates \cite{greiner_2003,*zwierlein_2003,*jochim_2003_science,*regal_2004}, the observation of reversible crossover to a degenerate Fermi gas \cite{bartenstein_2004,*regal_2003_nature}, measurements of collective excitation modes as well as pairing in a strongly interacting Fermi gas of atoms \cite{bartenstein_2004_2,*chin_2004_science}, and proofs of superfluidity in Fermi gases \cite{zwierlein_2005,*zwierlein_2006,*ku_2012,*wang_2019}.  

The crossover from a molecular Bose-Einstein condensate (BEC) to atomic Cooper pairs in the Bardeen-Cooper-Schrieffer state (namely, the BEC-BCS crossover) near a Feschbach resonance has been widely studied through several theoretical approaches, such as Quantum Monte Carlo methods \cite{astrakharchik_2004,*astrakharchik_2005,*jauregui_2007}, field theory \cite{duine_2004,*romans_2005}, and multi- or two-channel calculations \cite{bartenstein_2005,*duan_2006,*chin_2005,*wasak_2014}. However, as pointed in Refs. \cite{romans_2005,partridge_2005,liu_2019}, most of these theories fail to reproduce the measured closed-channel fraction and show considerable disagreement with available experimental data above the resonance \cite{partridge_2005,liu_2019}. The recent experiments of Ref. \cite{liu_2019} reinforce that even in very satisfactory matches (as the achieved in Refs. \cite{romans_2005, diehl_2006} by developing a functional integral formalism for atom and molecule field, or in Refs. \cite{chen_2005, liu_2019} within a two-channel pairing fluctuation theory) there is a considerable discrepancy between theory and experiments in the near-BCS regime. Here we present a simple two-channel model for two harmonically trapped atoms with finite-range interaction near a Feshbash resonance. We show that when including an effective trap which accounts not only for the optical trap and its geometry but also for the nontrivial many-body correlations the two-body model leads to handy general and near-resonance formulas for the binding energy and the closed-channel contribution, as well as to intuitive and accurate results.  

%%%%%%%%%%%%%%%%
%%%%%%%%%%%%%%%%
% Model and Results
%%%%%%%%%%%%%%%%
%%%%%%%%%%%%%%%%

The qualitative essence of the BEC-BCS crossover involves a continual change between a BEC of diatomic molecules (that in the case of a Fermi gas implies the emergence of a bosonic degree of freedom) and a BCS loosely correlated Cooper pairing state. This simple idea points to the need of considering diatomic molecules that are more and more weakly bound \cite{fermi_school_2006_book}. A simple model that enables such a pairing of atoms consists of two channels (open and closed) in which a two-body bound state can be created. The open-channel corresponds to two atoms while the closed-channel provides bare molecular states \cite{partridge_2005}. Then, the complete picture consists of a pair or dressed molecule in a superposition of the open- and closed-channel states,

%%%%%%%%%%%%%%%%
%%%%%%%%%%%%%%%%
\eq
%%%%%%%
\label{eq_dm_state}
%%%%%%%
\ket{\psi_{pair}} = \psi_m\, \ket{\text{closed}} + \psi_{aa}\, \ket{\text{open}} ,
\nonumber
%%%%%%%
\en 
%%%%%%%%%%%%%%%%
%%%%%%%%%%%%%%%%

\noindent with normalization $1 = \int \vert \psi_m \vert^2 d^3r  + \int \vert \psi_{aa} \vert^2 d^3r $ \cite{kohler_2006, chin_2010_review}. Therefore, $\int \vert \psi_m \vert^2 d^3r $ gives the probability of the pair to be in the closed-channel or the closed-channel population or fraction, usually denoted by $Z$ (here the simple idea is that each pair behaves as our modeled pair, then the number of pairs in the closed-channel satisfies $N_{c} = N_{pairs} \int \vert \psi_m \vert^2 d^3r  = (N/2) \int \vert \psi_m \vert^2 d^3r $, where $N_{pairs}$ is the number of pairs and $N$ the number of atoms, equivalently $\int \vert \psi_m \vert^2 d^3r  = 2 N_{c}/N = Z$ \cite{chen_2005, werner_2009}). Our previous qualitative formulation of the crossover requires $Z \sim 1$ deep in the BEC side and $Z \sim 0$ on the BCS side. Translating this into equations, the wave function of two trapped atoms with mass $m$ on an open channel supporting the threshold of the two-atom state and a closed-channel supporting a bound state $E_c$ magnetically tuned close to the threshold satisfies    

%%%%%%%%%%%%%%%%
%%%%%%%%%%%%%%%% 
\begin{eqnarray}
%%%%%%%
\label{eq_model}
%%%%%%% 
E \ket{\psi_{pair}} &=&  \left( - \frac{\hbar^2}{m} \nabla^2 + \frac{m \omega}{4} r^2+ \hat{v} \right) \ket{\psi_{pair}} 
%\nonumber \\
\\
%%%%%%%
\hat{v}&=& \left\{
\begin{array}{cc}
  - \frac{\hbar^2}{m} \, \left(
    \begin{array}{cc}
         q_o^2 & \Omega \\
         \Omega &  q_c^2 - \frac{m}{\hbar^2} \left( E_c + \mu B \right) 
    \end{array} 
    \right) & \text{for}\, r \leq r_0  \\
    & \\
    \left(
    \begin{array}{cc}
         0 & 0 \\
         0 & \infty 
    \end{array} 
    \right)& \text{for}\, r>r_0 , 
\end{array}
\right.
\nonumber
\end{eqnarray} 
%%%%%%%%%%%%%%%%
%%%%%%%%%%%%%%%%

\noindent where we consider spherical attractive potentials with range $r_0$ and depths $-\hbar^2 q_{o/c}^2/m$, a coupling between channels given by $\Omega$, a trap frequency denoted by $\omega$, and the Zeeman shift $\mu B$. To solve Eq.~\eqref{eq_model} one must introduce new superposition states, $\ket{+} = \cos\theta \,\ket{\text{open}} + \sin\theta \,\ket{\text{closed}}$ and $\ket{-} = -\sin\theta \,\ket{\text{open}} + \cos\theta \,\ket{\text{closed}}$, related to new dressed uncoupled channels \cite{chin_2005, wasak_2014}. The scattering length $a$ is obtained by solving the free-trap zero-energy scattering equation \cite{lieb_2000, giorgini_2008_review}, and can be rewritten in terms of the magnetic field as     

%%%%%%%%%%%%%%%%
%%%%%%%%%%%%%%%%
\eq
%%%%%%%
\label{eq_adeB}
%%%%%%%
\frac{a-r_0}{a_{bg}-r_0}=1+\frac{\Delta B}{B-B_{res}},
%%%%%%%
\en 
%%%%%%%%%%%%%%%%
%%%%%%%%%%%%%%%%

\noindent with $\Delta B$ being the resonance width and $B_{res}$ the resonance position. These quantities are given by $\Delta B=-\hbar^2 \gamma(a_{bg}-r_0)/m \mu$ and $B_{res} = - \hbar^2 \epsilon_c/ m \mu + \Delta B$, where $a_{bg}$ is the background scattering length, $\gamma = 2 q_c^2 \theta^2/r_0$ is the Feshbach coupling, and $\theta$ is the mixing angle of the dressed states \cite{chin_2005, wasak_2014}. The scattering length $a$ diverges when $B$ is tuned very close to the resonance. In this situation, known as the unitary limit, the interaction changes from attractive ($a>0$, molecular side -BEC) to repulsive ($a<0$, atom-atom side -BCS).

The energy of the two-body state obtained when solving Eq.~\eqref{eq_model} without restrictions is determined by

%%%%%%%%%%%%%%%%
%%%%%%%%%%%%%%%%
\eq
%%%%%%%
\label{eq_energy_complete}
%%%%%%%
\frac{\lambda}{\frac{D_{\lambda}(x_0)}{D_{\lambda-1}(x_0)}} =  
\cos\theta^2 \frac{\lambda_{+}}{\frac{f^{-}_{\lambda_+}(x_0)}{f^{+}_{\lambda_+ -1}(x_0)}}+
\sin\theta^2 \frac{\lambda_{-}}{\frac{f^{-}_{\lambda_-}(x_0)}{f^{+}_{\lambda_- -1}(x_0)}} ,
%%%%%%%
\en 
%%%%%%%%%%%%%%%%
%%%%%%%%%%%%%%%%

\noindent where $f^{\pm}_{\varsigma}(x) = D_{\varsigma}(x)\pm D_{\varsigma}(-x)$ with $D_{\varsigma}(x)$ being the Parabolic Cylinder functions \cite{abramowitz_stegun_1964_book, avakian_1987}, and $x_0 = \sqrt{m\omega/\hbar} r_0$. These functions depend on $\lambda = \epsilon - 1/2$ where $\epsilon = E/\hbar\omega$, $\lambda_{+} = \lambda +  \tilde{q}_o^2$, and $\lambda_{-} = \lambda +  \tilde{q}_c^2 - \epsilon_c - \mu B /\hbar \omega $. For the last definitions we used $\tilde{q}_{o/c}^2 = \hbar q_{o/c}^2/ m \omega$ and $\epsilon_c = E_c/\hbar\omega$. We also used the weak coupled channels conditions, i.e. $\Omega \ll q_o^2, q_c^2, \vert q_o^2 - q_c^2 \vert$ implying $\theta \ll 1$, which constitute an excellent approximation \cite{chin_2005, bouvrie_2017}. Notice that all the parameters are divided by the trap's characteristic length or energy. Taking into account several properties of the Parabolic Cylinder functions, assuming that the states are close to the threshold, and considering the experimental ranges of the involved quantities \footnote{Besides the conditions stressed in Ref. \cite{chin_2005}, the trap length must be large compared to the interaction range ($x_0 \ll 1$). In the Li case this requires $\omega \leq 2 \pi 10^7$ that (anticipating some results) corresponds to the condition $T_F \leq 500 \,\mu\text{K}$, which is satisfied in current experiments.}, Eq.~\eqref{eq_energy_complete} transforms into 

%\elo{Esto es muy importante. Las aproximaciones se basan en que la long de la trampa sea mayor que el rango de la interaccion $r_0$ que a lo sumo es 100 radios de Bohr. Para el caso del Li teniamos como tope una frecuencia de $2 \pi 10^7$ (para esta frecuencia la relacion es de $l_{\omega} = \sqrt{2 \hbar/ m \omega} \sim 11.6 r_0$). Este tope se corresponde con 500 $\mu$K que es 1000 veces el rango en el que se hacen los experimentos. Con esto nos quedamos mas que tranquilos. No obstante, si nos encontrasemos con temperaturas de Fermi mayores que esa simplemente habria que usar la formula completa y no la aproximacion.}

\begin{small}
%%%%%%%%%%%%%%%%
%%%%%%%%%%%%%%%%
\eq
%%%%%%%
\label{eq_energy_tg}
%%%%%%%
-\frac{\sqrt{2}\Gamma ( \frac{1-\lambda}{2} )}{\Gamma (-\frac{\lambda}{2} )} =  
\frac{\cos\theta^2\sqrt{\tilde{q}_{o}^2+\lambda}}{\tan ( \sqrt{\tilde{q}_{o}^2+\lambda}\, x_0 )}+
\frac{\sin\theta^2\sqrt{\bar{q}_{c}^2 +\lambda}}{\tan ( \sqrt{\bar{q}_{c}^2+\lambda} \,x_0 )} ,
%%%%%%%
\en 
%%%%%%%%%%%%%%%%
%%%%%%%%%%%%%%%%
\end{small}
  
\noindent where $\bar{q}_{c}^2 = \tilde{q}_{c}^2 - \epsilon_c - \mu B/\hbar \omega$. If $\lambda <0$, the well known expansion  $ \Gamma(z+1/2)/\Gamma(z) = \sqrt{z} ( 1- 1/8z + \cdots )$ in the left side of the above equation leads to Eq. (12) of Ref. \cite{chin_2005} as a first order approximation to the energy relative to the ground state of the trap. Since we are particularly interested in the near-resonance crossover coinciding with $\lambda \sim 0$, we need to keep all the physics hidden behind this term. The left side of Eq.~\eqref{eq_energy_tg} adds mostly the trap effect (it is the same term that arises when solving a single channel delta-type interaction with a trap \cite{avakian_1987, busch_1998}), while the right side contains the two-channel free-trap physics.

Now we focus on the derivation of handy formulas for the binding energy of the molecules and for the closed-channel fraction, both accesible quantities in current experiments. Regarding the conditions mentioned earlier and following similar calculations as those presented by C. Chin in Ref. \cite{chin_2005}, Eq.~\eqref{eq_energy_tg} reduces to

%%%%%%%%%%%%%%%%
%%%%%%%%%%%%%%%%
\eq
%%%%%%%
\label{eq_energy_simpler}
%%%%%%%
( \frac{\sqrt{2}\Gamma ( \frac{1-\lambda}{2} )}{\Gamma ( -\frac{\lambda}{2} )} + \frac{1}{x_0 - \tilde{a}_{bg}} ) ( \epsilon_c + \frac{\mu B}{\hbar \omega} - \lambda ) = \tilde{\gamma} ,
%%%%%%%
\en 
%%%%%%%%%%%%%%%%
%%%%%%%%%%%%%%%%       

\noindent with $\tilde{a}_{bg} = \sqrt{m\omega/\hbar} \,a_{bg}$ and $\tilde{\gamma} = \gamma/ (m\omega/\hbar)^{3/2}$. When the coupling between channels is absent ($\gamma =0$) Eq.~\eqref{eq_energy_simpler} implies $\lambda = \epsilon_c + \mu B/\hbar \omega$ and $\sqrt{2}\Gamma (1/2-\lambda/2)/\Gamma(-\lambda/2) = 1/(\tilde{a}_{bg}-x_0)$. The former corresponds to the bound state in the closed-channel while the latter resembles the results obtained when considering a single channel in a trap \citep{avakian_1987,busch_1998}. 

The closed-channel fraction $Z$ can be obtained by direct integration of the closed-channel wave function $Z = \int \vert \psi_m \vert^2 d^3r $ thus requiring numerical integration, or as the derivative of the energy on $\epsilon_c$, i.e. $Z = \partial \lambda/ \partial \epsilon_c$, due to the Hellman-Feynmann theorem \cite{chin_2010_review, chin_2005, werner_2009}. Although both procedures provide the same result, the second one leads directly to 

\begin{small}
%%%%%%%%%%%%%%%%
%%%%%%%%%%%%%%%%
\eq
%%%%%%%
\label{eq_Z_simpler}
%%%%%%%
Z = \frac{2 \tilde{\gamma}}{2\tilde{\gamma} + ( \epsilon_c + \frac{\mu B}{\hbar \omega} - \lambda )^2 \frac{\sqrt{2}\Gamma ( \frac{1-\lambda}{2} )}{\Gamma (-\frac{\lambda}{2} )} \left\lbrace \Psi ( \frac{1-\lambda}{2} ) - \Psi ( -\frac{\lambda}{2} ) \right\rbrace} ,
%%%%%%%
\en 
%%%%%%%%%%%%%%%%
%%%%%%%%%%%%%%%%     
\end{small}

\noindent where $\Psi(z)$ denotes the Digamma function \citep{abramowitz_stegun_1964_book}. For a given magnetic field, one must first solve Eq.~\eqref{eq_energy_simpler} to obtain the ground state energy and then insert it in Eq.~\eqref{eq_Z_simpler}. It is possible to obtain even simpler near-resonance expressions. Expanding Eq.~\eqref{eq_energy_simpler} for small $\lambda$ and using Eq.~\ref{eq_adeB}, the dependence of the molecular binding energy on the scattering length and magnetic field reads      

%%%%%%%%%%%%%%%%
%%%%%%%%%%%%%%%%
\eq
%%%%%%%
\label{eq_energy_B}
%%%%%%%
\frac{\sqrt{2}\Gamma ( \frac{1-\lambda}{2} )}{\Gamma ( -\frac{\lambda}{2} )} = \frac{1}{\tilde{a} - x_0} = \frac{\mu (B-B_{res})}{\hbar \omega \tilde{\gamma} (\tilde{a}_{bg} -x_0)^2},
%%%%%%%
\en 
%%%%%%%%%%%%%%%%
%%%%%%%%%%%%%%%%

\noindent where $\tilde{a} = \sqrt{m\omega/\hbar} \,a$. Since the characteristic length of the trap is larger than the range of the interaction, the obtained dependence of the molecular binding energy on the scattering length is essentially the same obtained for a delta potential plus a correction due to the interaction range. %\elo{Esto es muy importante. Las aproximaciones se basan en que la long de la trampa sea mayor que el rango de la interaccion $r_0$. Para el caso del Li teniamos como tope una frecuencia de $2 \pi 10^7$ (para esta frecuencia la relacion es de $l_{\omega} = \sqrt{2 \hbar/ m \omega} \sim 11.6 r_0$). Este tope se corresponde con 500 $\mu$K que es 1000 veces el rango en el que se hacen los experimentos. Con esto nos quedamos mas que tranquilos. No obstante, si nos encontrasemos con temperaturas de Fermi mayores que esa simplemente habria que usar la formula completa y no la aproximacion.}

Although the free-trap two-body theory predicts that the closed-channel fraction vanishes when the resonance is reached (due to the absence of a two-body bound state for $a<0$ in free space), the experimental evidence shows that it continues smoothly across the resonance \cite{chin_2005, falco_2005, partridge_2005, chen_2005, liu_2019}. Using the near-resonance approximation of Eq.~\eqref{eq_energy_B} given by $\lambda = \mu (B-B_{res})/\hbar \omega ( 1+ \sqrt{\pi/2} \tilde{\gamma} (\tilde{a}_{bg} -x_0)^2 )$ in a first order expansion of Eq.~\eqref{eq_Z_simpler}, it is straightforward to see that the non-vanishing closed-channel contribution in the resonance is

%%%%%%%%%%%%%%%%
%%%%%%%%%%%%%%%%
\eq
%%%%%%%
\label{eq_Z_res}
%%%%%%%
Z_{res}  = \frac{1}{1+ \sqrt{\frac{\pi \hbar}{2 m \omega}} \gamma (a_{bg} -r_0)^2} . 
%%%%%%%
\en 
%%%%%%%%%%%%%%%%
%%%%%%%%%%%%%%%% 

Now the naive idea is that the trap in our model is an effective trap accounting not only for the optical trap but also for the nontrivial many-body correlations. The effective trap frequency is given by $\hbar \omega_{eff} = k_B T_F = \hbar \bar{\omega} \,(3 N)^{1/3}$, where $k_B$ is the Boltzmann constant, $T_F$ the Fermi temperature, and $\bar{\omega}$ the geometric mean of the three frequencies of the external trap \cite{giorgini_2008_review, werner_2009}. In other words, the characteristic energy of the trap is given by the Fermi temperature of a harmonically trapped ideal Fermi gas, and its characteristic length is comparable to the interparticle spacing. The effective trap takes into account the geometry of the optical trap (present in $\bar{\omega}$) as well as the number of atoms $N$. For vanishing $\bar{\omega} \,(3 N)^{1/3}$ the trap is absent and $Z_{res}$ goes to zero recovering the results of the free-trap model of Ref. \cite{chin_2005}. Numerical integration of the wave function gives $\langle r \rangle = a/2$ for fields below the resonance width, in consonance with the results of the regularized delta and two-channel free models \cite{giorgini_2008_review, chin_2005}. Deeply into the BEC side the molecules behave as point-like composite bosons unaffected by the effective trap, while the pair size grows towards the resonance. When the available space defined by the effective trap is large compared to the size of the pairs (BEC side), the trap enhances the closed-channel fraction in line with the intuitive notion that the trap forces the pairs to be in a molecular state. When the size of the molecules begins to be comparable to the effective trap's characteristic length (BCS side), the trap acts as a buffer for the closed-channel contribution. Therefore, the effective trap simulates all the remaining fermions. It simulates the insufficient physical space favoring the interaction between pairs and provides a mechanism leading to Pauli blocking because the unavailability of enough space in the real space is related to the unavailability of sufficient space in the state space \cite{chudzicki_2010, cuestas_2020}.

%%%%%%%%%%%%%%%%
%%%%%%%%%%%%%%%%
% Comparison with experiments and theory
%%%%%%%%%%%%%%%%
%%%%%%%%%%%%%%%%

Using $\omega = \omega_{eff} = k_B T_F/ \hbar$ in the expansion of Eq.~\eqref{eq_Z_res} we obtain

%%%%%%%%%%%%%%%%
%%%%%%%%%%%%%%%%
\eq
%%%%%%%
\label{eq_Z_res_TF}
%%%%%%%
Z_{res}  = \frac{\sqrt{\frac{2 k_B m}{\pi \hbar^2}}}{\gamma (a_{bg} -r_0)^2} \sqrt{T_F}, 
%%%%%%%
\en 
%%%%%%%%%%%%%%%%
%%%%%%%%%%%%%%%% 

\noindent in agreement with the dependency of $Z$ on $T_F$ at unitarity predicted in Refs. \cite{chen_2005, zhang_2009, werner_2009} within different many-body approaches. In what follows we contrast our results with the available experimental data for the closed-channel fraction measured in a $^6$Li Fermi gas when crossing the so called $^6$Li broad resonance \cite{partridge_2005, liu_2019}, and with the theoretical results of Ref. \cite{chen_2005}. The corresponding parameters are given by $r_0 = 29.9\, a_0$, $B_{res} = 834.15$ G, $\Delta B = 300$ G, $a_{bg} = -1405 \, a_0$, $\mu = 2.0 \,\mu_B$, and $\gamma^{-1/3} = 101 \,a_0$, where $a_0$ and $\mu_B$ denote the Bohr radius and magneton respectively \cite{chin_2005, chin_2010_review}.

\begin{figure}[t]
%%%%%%%%
\includegraphics[width=\columnwidth]{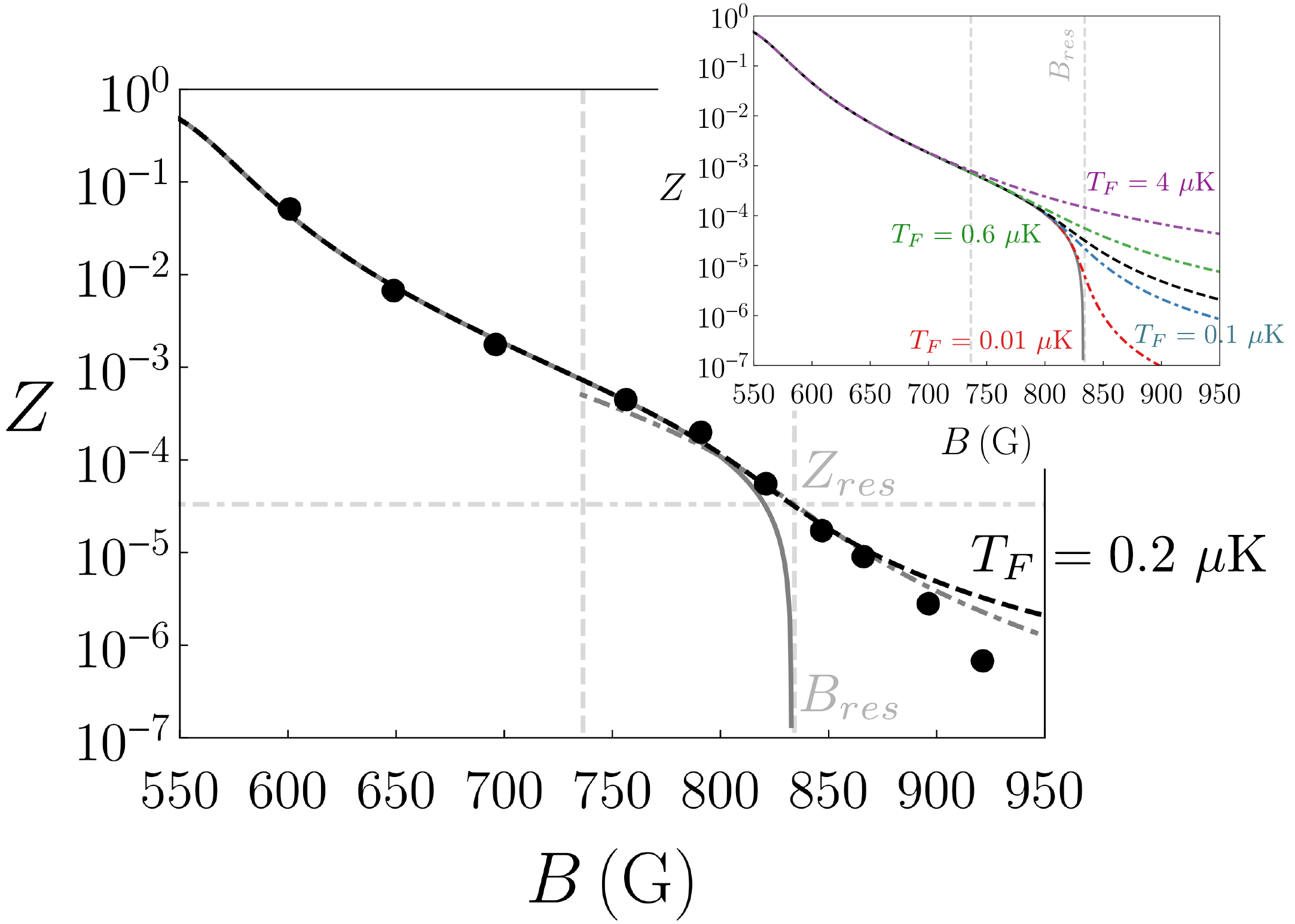}
%%%%%%%%
\caption{Closed-channel fraction $Z$ vs. magnetic field $B$. The points are the experimental data of Ref. \cite{partridge_2005}, whose size reflects the uncertainty in $Z$. The closed-channel fraction obtained with the free-trap model of Ref. \cite{chin_2005} is depicted in gray solid line while our results are depicted in black dashed line -see Eq.~\eqref{eq_Z_simpler}. The approximation calculated using Eq.~\ref{eq_energy_B} is shown in gray dot-dashed line. The horizontal lightgray dot-dashed line gives $Z_{res}$. Vertical lightgray dashed lines indicate the $B_{res}$ value and the typical BEC-BCS Crossover regime $|a|>3000\, a_0$. Notice that the 920 G point is identified in Ref. \cite{partridge_2005} as presenting experimental issues. The inset depicts the obtained $Z$ for several $T_F$ values.}
\label{fig_Z_partridge}
%%%%%%%%
\end{figure}

Figure~\ref{fig_Z_partridge} shows the closed-channel fraction $Z$ for magnetic fields between 600 and 950 G. The points are the experimental data taken from Ref. \cite{partridge_2005}, whose size indicates the uncertainty in $Z$. The gray solid line is the calculated $Z$ within the free-trap two-body model of Ref. \cite{chin_2005}. The black dashed line are the results obtained via Eq.~\eqref{eq_Z_simpler}, while the gray dot-dashed lines are obtained using Eq.~\ref{eq_energy_B} in Eq.~\ref{eq_Z_simpler}, both with $T_F = 0.2 \,\mu\text{K}$ (in Ref. \cite{partridge_2005} $T_F$ ranges between $0.2$ and $0.6 \,\mu\text{K}$). The trap-free model matches the data below the resonance but fails near and above the resonance. Our model leads to the same values than the free model below the resonance and shows good agreement near and above the resonance. To show that the free-trap results are recovered when $T_F$ is small enough, the inset of the figure shows the obtained $Z$ for different values of $T_F$. %In Ref. \cite{partridge_2005} the 920 G point is identified as presenting experimental issues, which can be the reason for its deviation.

% For $T_F = 0.2 \,\mu\text{K}$ we obtain $Z_{res} = 3.327 \,10^{-5}$ (horizontal lightgray dashed line).

% PRL chen 2005 dice, respecto de la Fig.1; "However, for the continuous curve plotted in the main figure, we use $T_F = 0.2 \,\mu\text{K}$ which better represents the values in the BCS regime [13]. In the inset we used $T_F = 0.4 \,\mu\text{K}$, since this best reflects the average value over the entire range of data points." 

%%%%%%%%%%%%%%%
%%%%%%%%%%%%%%%
%%%%%%%%%%%%%%%
%%%%%%%%%%%%%%%

\begin{figure}[ht]
%%%%%%%%
\includegraphics[width=\columnwidth]{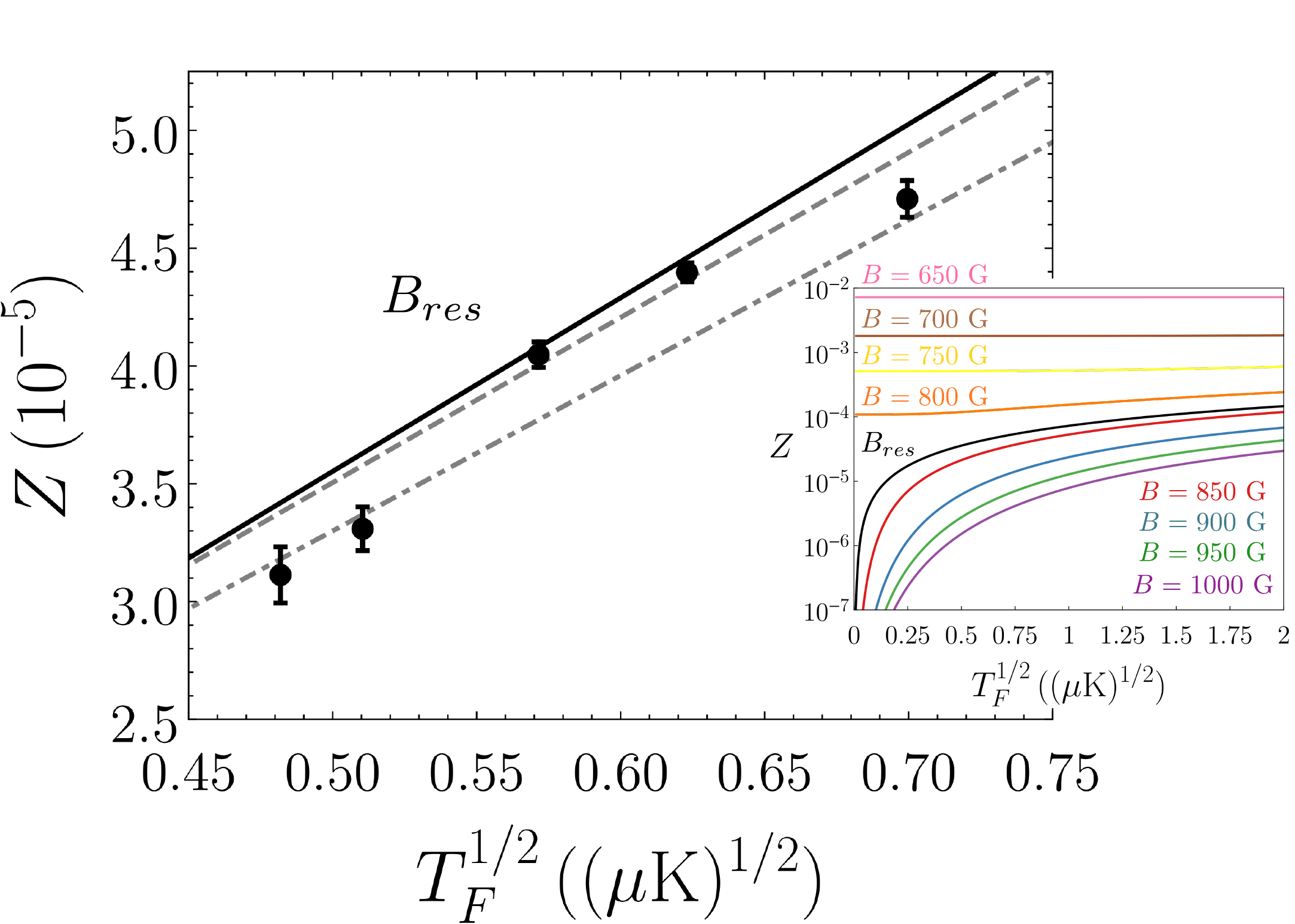}
%%%%%%%%
\caption{Closed-channel fraction $Z$ at unitarity vs. $\sqrt{T_F}$. The points are the experimental data extracted from Ref. \cite{liu_2019}. Our results are the black solid line. The gray dashed line and gray dot-dashed line represent the experimental fitting and the theoretical curve reported in Ref. \cite{liu_2019} respectively. The inset presents the behavior of $Z$ for several magnetic fields in qualitative agreement with the results of Ref. \cite{chen_2005} (we obtain a larger $Z$ on the BCS side).}
\label{fig_Z_liu_unitarity}
%%%%%%%%
\end{figure}

\begin{figure}[t]
%%%%%%%%
\includegraphics[width=\columnwidth]{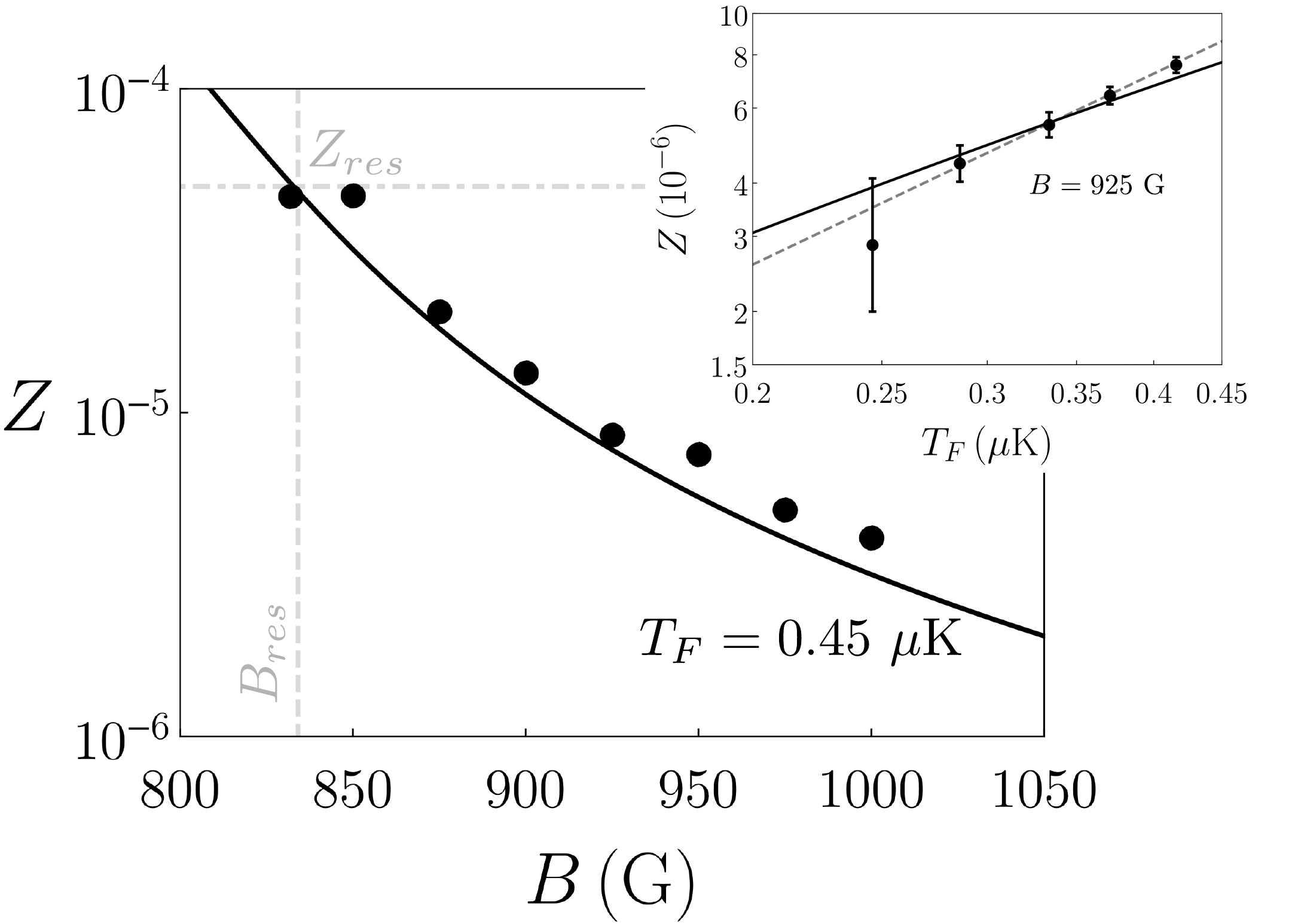}
%%%%%%%%
\caption{Closed-channel fraction $Z$ vs. magnetic field $B$ on the near-BCS side. The points are the experimental data extracted from Ref. \cite{liu_2019}, whose size includes experimental errors. Our results are depicted in black solid line. The horizontal and vertical lightgray dashed lines give $Z_{res}$ and $B_{res}$ respectively. The inset (loglog scale) shows the obtained $Z$ vs. $T_F$ for $B=925\,\text{G}$ together with the power law datafit reported in Ref. \cite{liu_2019}.}
\label{fig_Z_liu_BCS}
%%%%%%%%
\end{figure}

Figure \ref{fig_Z_liu_unitarity} depicts the dependency of the the closed-channel fraction on $T_F$ at unitarity. Our results (black line) are in close agreement with the experimental data (dots) and the datafit (gray dashed line) presented in Ref. \cite{liu_2019}. We obtain $Z = 0.074 \sqrt{T_F}$ at unitarity (see Eq.~\eqref{eq_Z_res_TF}), while the experimental and theoretical proportionality constant reported in Ref. \cite{liu_2019} are $0.074(12)$ and $0.066$ (gray dot-dashed line) respectively. To show that the obtained $Z$ is in qualitative agreement with previous theoretical results, the inset of Fig. \ref{fig_Z_liu_unitarity} depicts its behavior for several magnetic fields; at unitarity $Z$ goes as $\sqrt{T_F}$, on the BEC side it is less sensitive on $T_F$, and on the BCS side presents a higher power law \cite{chen_2005}. Notice that the obtained $Z$ is larger than the one calculated in Ref. \cite{chen_2005} on the BCS side, where a considerable disagreement between experiments and theory has been repeatedly observed \cite{partridge_2005, romans_2005, liu_2019}. Figure \ref{fig_Z_liu_BCS} shows the agreement between our results (black line) and the experimental data of Ref. \cite{liu_2019} (dots) for fields above resonance and $T_F = 0.45\,\mu\text{K}$. Finally, the obtained agreement with measurements and the power law datafit (gray dashed line) reported in Ref. \cite{liu_2019} for $B=925\,\text{G}$ is shown in the inset of Fig. \ref{fig_Z_liu_BCS}.  

%%%%%%%%%%%%%%%
%%%%%%%%%%%%%%%
%%%%%%%%%%%%%%%
%%%%%%%%%%%%%%%

%%%%%%%%%%%%%%%
%%%%%%%%%%%%%%%
%%%%%%%%%%%%%%%
%%%%%%%%%%%%%%%

% A threshold $T_F$ can be defined by requiring $Z_{res} = 1/2$ in Eq.~\ref{eq_Z_res}, $T_F^{th}  = \pi \hbar^2 \gamma^2 (a_{bg} -r_0)^4 / 2 k_B m $. For frequencies below (above) $T_F^{th}$, the resonance is open- (closed-) channel dominated, corresponding to $Z_{res}$ below (above) $1/2$. For the broad resonance of Li, $T_F^{th} = 45.17 \,\text{K}$. 

% From Eq.~\eqref{eq_Z_res} a relevant quantity arises, $ \chi_{res} = \sqrt{\hbar m/k_B T_F} \mu \Delta B (r_0 - a_{bg})$, which can be useful when comparing resonances for a fixed $T_F$: larger and positives $\chi_{res}$ leads to smaller $Z_{res}$. 
% Esta relevant quantity fue reemplazada por el $\alpha$ de la proporcionalidad.

%%%%%%%%%%%%%%%%
%%%%%%%%%%%%%%%%
% Summary/ conclusions
%%%%%%%%%%%%%%%%
%%%%%%%%%%%%%%%%

In conclusion, by adding a trap with characteristic energy given by the Fermi temperature $T_F$ to a simple two-channel two-body model for Feshbach resonances, we were able to reproduce the measured closed-channel fraction across the BEC-BCS crossover and into the BCS regime of a $^6$Li atomic Fermi gas \cite{partridge_2005, liu_2019}. We derived general and near-resonance simple formulas which show remarkable agreement with measurements and previous theoretical results \cite{chen_2005, zhang_2009, werner_2009}. We obtained the expected dependency $Z \propto \sqrt{T_F}$ at unitarity, with a proportionality constant in close agreement with experiments \cite{liu_2019}. Our results are also in agreement with recent measurements of the $Z$ dependency on $T_F$ on the near-BCS side, where a significant experiment-theory discrepancy has been reported \cite{partridge_2005, romans_2005, liu_2019}.

The effective trap accounts not only for the optical trap and its geometry but also for the nontrivial many-body fermionic correlations. Our naive picture is that all the remaining fermions and their correlations act as a trap supporting a bound state for the pair which leads, in turn, to a non-vanishing closed-channel fraction for fields above resonance. We would like to stress the simplicity of the model as well as its intuitive and accurate results. The effective trap simulates the remaining fermions because it emulates the insufficient physical space that favors the interaction between pairs and provides a mechanism for Pauli blocking \cite{chudzicki_2010, cuestas_2020}. This paves the way for studying the many-body unitarity physics by adding the exchange interactions within the composite boson ansatz \cite{leggett_2001,*combescot_2001,*bouvrie_2019}, whose construction of the many-particle state relies upon the availability of an accurate pairing model.   

% It is important to notice that our model can be applied to other elements.

% We would like to finish by suggesting that including this accurate two-body model in others approaches (as in the trial functions used in Montecarlo studies \cite{astrakharchik_2004,*astrakharchik_2005,*jauregui_2007}, or in the effective quantum field theory for Feshbach-resonant interactions \cite{duine_2004,*romans_2005}) may imply a considerable gain in the understanding of the many-body interacting quantum system, which is constructed upon a complete insight of the microscopic two-body physics.

%%%%%%%%%%%%%%%%
%%%%%%%%%%%%%%%%
% Acknowledgements
%%%%%%%%%%%%%%%%
%%%%%%%%%%%%%%%%

We are grateful to P. A. Bouvrie for introducing us in these questions. We acknowledge funding from grant PICT-BID 2017-2583 from ANPCyT and grant GRFT-2018 MINCYT-Córdoba, as well as financial support from SeCyT-UNC and CONICET. We would like to thank the reviewers because their criticism and comments helped us to substantially improve the present work. We would like to make a last unusual acknowledge: to C. Chin for his generosity regarding knowledge, because his arXiv-published work, Ref. \cite{chin_2005}, inspired the present discussions.

%%%%%%%%%%%%%%%%
%%%%%%%%%%%%%%%%

\bibliography{two_channels}
\bibliographystyle{h-physrev5}

%%%%%%%%%%%%%%%%
%%%%%%%%%%%%%%%%

\end{document}